# Hanny's Voorwerp and the Antikythera Mechanism – similarities, differences and insights.

**Michael A. Garrett**[1]

*ASTRON, Netherlands Institute for Radio Astronomy, Postbus 2, 7990 AA, Dwingeloo, The Netherlands*
*E-mail:* garrett@astron.nl

I present some insights into Hanny's Voorwerp and the Antikythera mechanism – contrasting their similarities and differences. They are both excellent examples of serendipitous discoveries in which human curiosity and perseverance have played an important role. Both objects have captured the imagination of the general public, and their discovery was only made possible via the introduction of new technologies. One major difference is that there is only one Antikythera device but there are now many Voorwerpen or "voorwerpjes", as they are more commonly known. The study of a collection of objects, as is common in astronomy, greatly aids our understanding of cosmic phenomena. In the case of the voorwepjes, we now know that such systems are to be identified with obscured galaxies or Active Galactic Nuclei (AGN) that appear to have recently and indeed rapidly turned off.

Clearly, the discovery of more examples of devices similar to the Antikythera mechanism would have a significant affect in advancing our understanding of this object and the people that constructed it. Thus far, surveys of the site of the Antikythera wreck are incomplete and non-systematic. Like radio astronomy and other progressive fields, technological advances proceed exponentially in terms of capacity and capability. Recent advances in diving technology are no exception to this rule. It is almost 40 years ago that Jacques Cousteau led the last adhoc survey of the Antikythera wreck – the time has surely come to revisit the site and conduct a proper scientific and systematic survey. The deepest areas of the site are so far completely unexplored while it is known that some artefacts did fall into this area during the original excavation.

During this workshop, I called for a return to the site using the most modern technologies e.g. AUVs, ROVs, magnetometers and metal detectors. The hull of the wreck is known to be in good condition, buried under several feet of sediment. Any cargo remaining buried in the ship would be very well preserved. In addition, the ship carrying the Antikythera device may have been part of a larger convoy - sonar equipment should be able to detect other possible wrecks in the vicinity of the original site. The response to my call from experts on the Antikythera mechanism was initially subdued but at the time of writing this article, I learned that a new survey, undertaken by the Wood Hole Oceanographic Institute (USA) is underway at the Antikythera site. We wait with interest, and not a little anticipation for the first results to appear!

---

[1] Speaker

## 1. Introduction

Hanny's Voorwerp[2] (SDSS J094103.80+344334.2) is an irregular, highly ionised gas cloud located ∼ 25 kpc to the southeast of the massive disk galaxy IC 2497 [1]. Paradoxically, there is no clear evidence of an ionising source in the immediate proximity of the galaxy or the associated nebulosity. This galaxy scale gas cloud was first detected by Dutch school teacher, and citizen scientist, Hanny van Arkel. Considered to be one of the most interesting discoveries to emerge from the Galaxy-Zoo project, the Voorwerp may be the first example of a quasar light echo [1]. Indeed X-ray observations [2] interpreted in this context conclude that the galaxy's central engine has decreased its radiative output by at least 2 orders of magnitude within the last 70,000 years. Observations of neutral hydrogen in IC2497, using the Westerbork Synthesis Telescope (WSRT), show that the Voorwerp (See Figure 1) is part of a huge reservoir of gas that surrounds IC2497 [3].

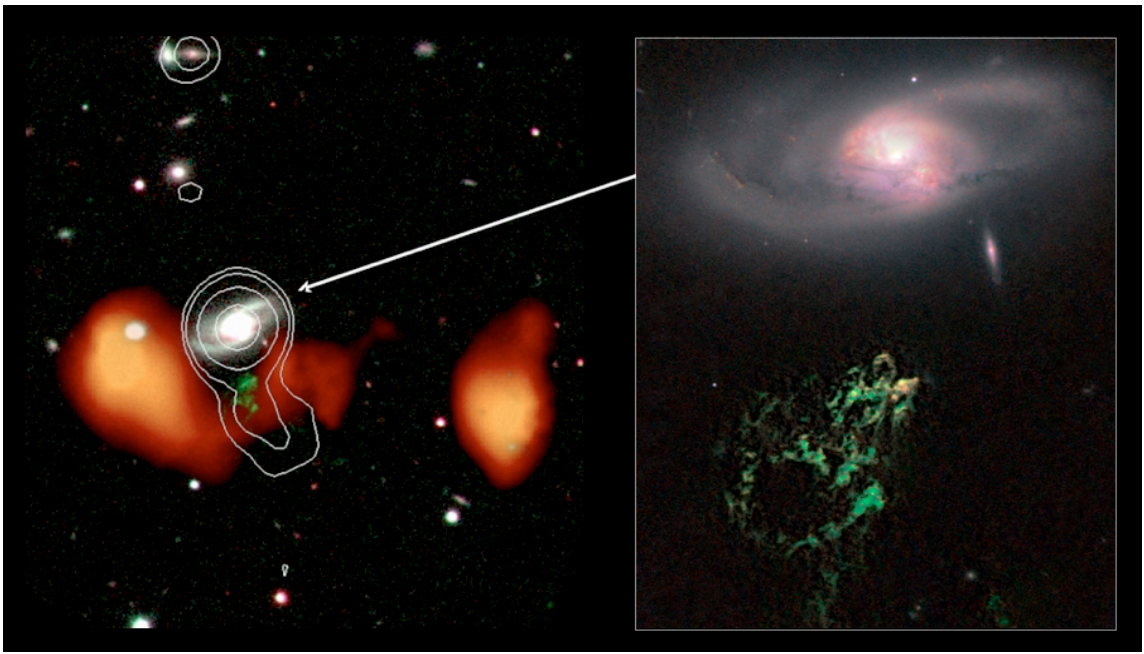

*Figure 1: Radio continuum (white contours) and neutral hydrogen emission (orange colours) are superimposed on an optical image of the field associated with IC2497 and Hanny's Voorwerp [3]. The recent HST images (boxed right) are also presented [5].*

High-resolution European VLBI Network (EVN) observations detect compact radio emission in the centre of the galaxy, strongly suggesting that some AGN activity is on-going. Intermediate resolution MERLIN observations show the bulk of the radio emission to be extended in nature on sub-kpc scales, and this, together with the galaxies FIR luminosity argue that the AGN is conincident with a nuclear starburst with a total star formation rate (assuming a Salpeter IMF) of ~ 70 $M_\odot$/yr [4]. In addition, to detecting neutral hydrogen in emission, the WSRT also detects it in absorption towards the central radio source. On larger scales, the WSRT radio continuum observations show faint, one-sided, large-scale emission (interpreted as a radio jet) pointing exactly in the direction of Hanny's Voorwerp [3], perpendicular to the galaxies major axis. An alternative interpretation to the light-echo hypothesis is that the radiation cone of the AGN is highly obscured along the observers line-of-sight [6]. However, superbly detailed HST images of the field [5], together with the non-detections of hard x-ray emission from the Suzaku satellite of IC2497 [2] are consistent with the light-echo hypothesis.

---

[2] "Voorwerp" is a Dutch word meaning "Object".

The Antikythera mechanism is a 2000-year-old mechanical device designed to calculate the position of the sun and moon, as well as to calculate the dates of solar and lunar eclipses. It is likely that it could also calculate the positions of the inner planets (see these proceedings, and in particular Seiradakis 2012). The device was discovered as part of an excavation of the seabed that started serendipitously in October 1900, when sponge divers discovered an ancient shipwreck, off the Greek island of Antikythera (see Figure 2). Sheltering from a storm, they discovered that the seabed was littered with priceless statues and ancient relics of Roman Greece, dating back to before 100 BC. Together with the Greek Education Ministry and Hellenic Navy, the sponge divers salvaged numerous artifacts from the waters over the next year. In May 1902, pieces of bronze recovered from the expedition, were recognised by Spyridon Stais at the Greek National Archaeological Museum in Athens, as a set of intricate gears and wheels. Recent X-ray tomography of the parts has revealed the so-called Antikythera mechanism to be a complex device that contains 30 bronze gears and about 224 teeth. Dating back to the time of Hipparchos, the complexity of the device is truly impressive, as is the astronomical model that governs its inner workings. Certainly there seems to be no device of comparable complexity built for at least a millennium after the Antikythera mechanism was constructed.

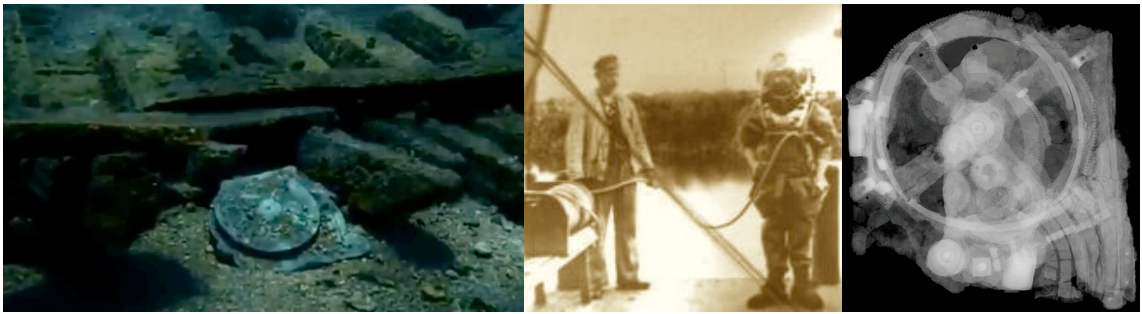

*Figure 2: The Antikythera mechanism as it might have appeared on the surface of the seabed (left) on its discovery more than 110 years ago. Middle: Antikythera diver suited up and ready to dive (ca. 1902). Right: X-ray tomography of the mechanism showing the complexity of its inner workings.*

In this paper, I compare and contrast both the discovery and study of Hanny's Voorwerp and the Antikythera mechanism. There are many similarities and differences – hopefully the analysis provides some insights into the study of cosmic phenomena and objects of antiquity.

## 2. Hanny's Voorwerp & the Antikythera mechanism

2.1 Similarities

There are many similarities between the discovery and study of Hanny's Voorwerp and the Antikythera mechanism. First of all, they were both very much serendipitous discoveries. Hanny van Arkel, having viewed thousands of galaxy images, realised that there was something unusual with IC2497. Instead of dismissing the "blue stuff" in the image, Hanny posted the image on the GalaxyZoo forum and reported her observations – "What's this blue stuff below? Anyone?". With a detection in only one of the SDSS filters, others might surely have concluded that this was a artefact of the image. Hanny did not. Similarly, Spyridon Stais (former Greek Minister of Education), visiting the Archaeological Museum noticed by chance that some pieces of bronze antiquities encrusted with rock revealed under closer inspection gear-like structures. These rusty bronze parts might have easily been dismissed as uninteresting compared to the many beautiful bronze and marble statues that surrounded them. However, Spyridon, piecing together several of the bronze parts together and deciphering the ancient inscriptions realized that this was an important discovery – the Antikythera mechanism had been found.

Both the Voorwerp and the Antikythera mechanism have also captured the imagination of not only the scientific community but the public at large– both objects attract an unusual amount of press attention (see Nikoli & Seiradakis 2012, these proceedings). In addition, both discoveries arose via the use of new technologies – in the case of the Voorwerp, the introduction of crowd-sourcing via the GalaxyZoo Citizen Scientist effort using the internet, in the case of the Antikythera mechanism, via the recent invention of canvas diving suits equipped with copper helmets and air breathing lines – this allowed divers to go deeper and stay submerged longer than ever before.

Scientists are driven to understand what these objects are and how they work. In the case of the Voorwerp, we also want to know what it tells us about the nature of the universe, our place in it, and perhaps also something about the creator (should we believe in one). For the Antikythera mechanism, we are interested in what it tells us about history, human development and the nature of the people that constructed it. For individual scientists, the element of competition is a driving force in the study and analysis of both objects – every scientist wants to be the first to fully understand the inner workings of these systems, and before anybody else! This competitive approach was very much in evidence at the workshop. However, a more cooperative approach to advancing science was also apparent (see Jauncey 2012, these proceedings).

2.2 Differences

There are also some differences regarding the study of Hanny's Voorwerp and the Antikythera mechanism. An obvious advantage to the study of the Voorwerp is that there are now other examples of this phenoema (see Fig 3) – new but typically smaller scale examples known as "voorwerpjes[3]". The study of about 20 of these voorwerpjes (see Figure 3) has led to the conclusion that these systems are associated with either obscured but active AGN or AGN that have recently turned off, and whose light we only see via a light-echo reflection, similar to the Voorwerp itself.

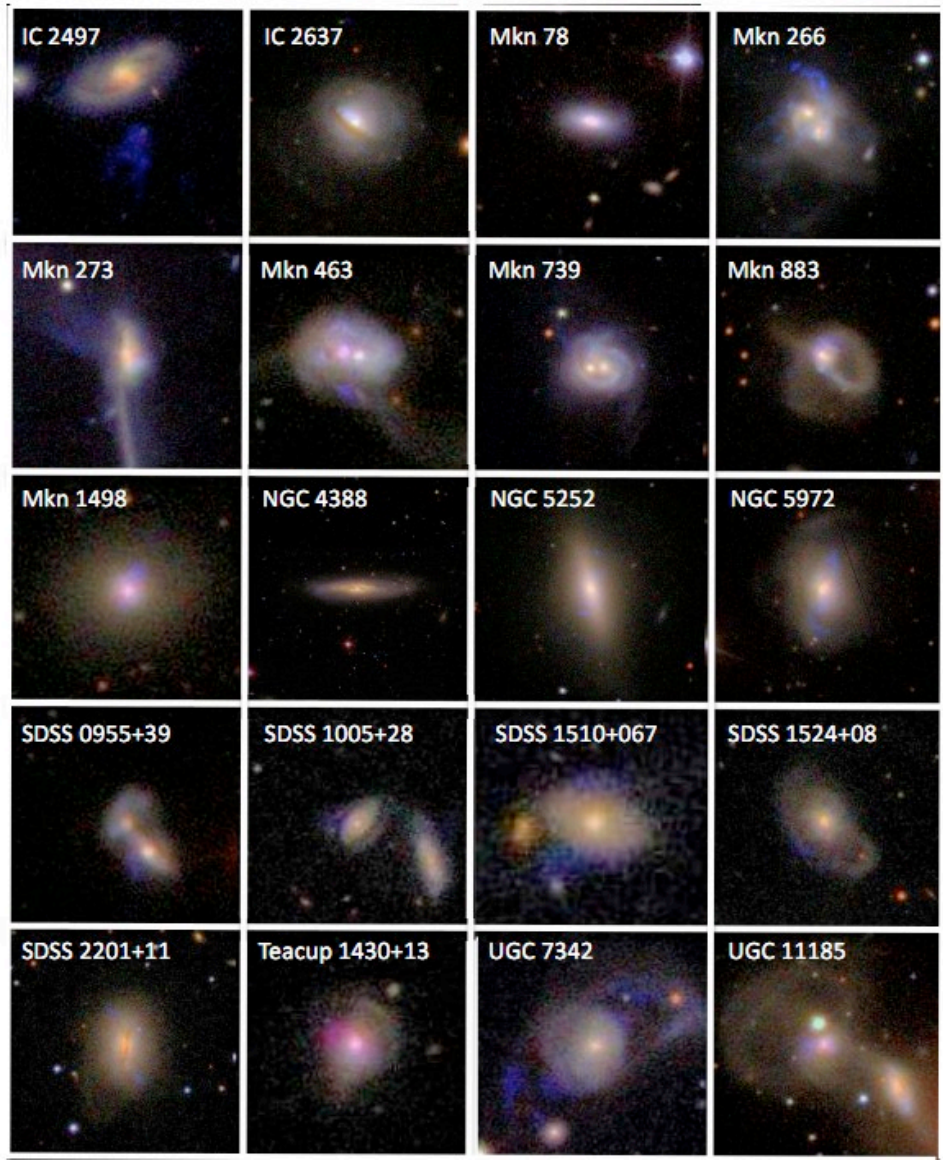

*Figure 3: Examples of other ionised nebula, similar in nature to Hanny's Voorwerp but typically smaller in scale (See Keel et al. [7]).*

The Antikythera mechanism, on the other hand, is the only surviving example of a technology that was lost through the ages and only recently re-discovered. Unfortunately, there is only one example of the device to study, although there were probably several, perhaps many such devices in circulation 2000 years ago. This circumstance leads us to consider whether a new search for more examples is not warranted (see Section 3).

---

[3] "Voorwerpjes" is the diminutive form of the Dutch word Voorwerpen)

**3. Returning to the Antikythera wreck**

There have been two expeditions to the Antikythera site – the first in 1900-01 and a second by Jacques Cousteau in 1976 (see Tsavliris 2012, these proceedings). Neither of these expeditions were thorough or systematic, and part of the site is completely unexplored (see later - Figure 5). Several divers from the 1900 expedition recall at least one large artefact (a bronze horse head) falling into the unexplored region of depth ~ 200 metres. One theme of the workshop explored by Ekers (see these proceedings) was the exponential rise in capability in fields such as radio astronomy, and in particular the sensitivity and speed of the instruments being used to survey the sky. This exponential increase is also seen in other progressive areas, one pertinent area of interest to the following discussion is diving – during the original 1901 expedition, divers could only remain under the water for a few minutes at most.

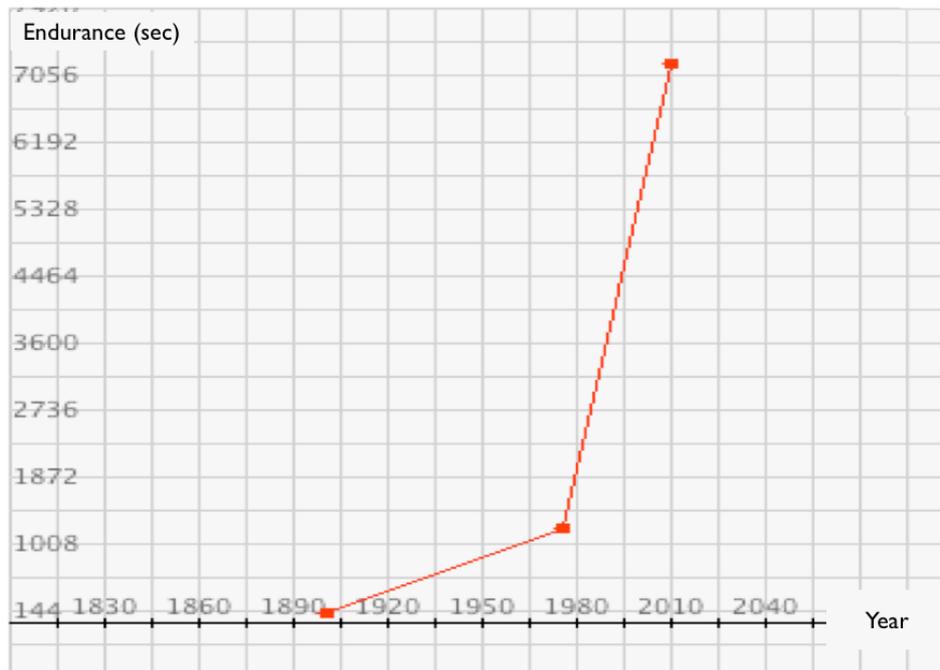

*Figure 4: Another example of exponential progress in a field other than radio astronomy – deep-sea divers underwater endurance times versus year.*

Nowadays, divers can remain at such depths for several hours using re-breathing apparatus. Figure 4 shows this exponential rise in underwater endurance times over the last 110 years.

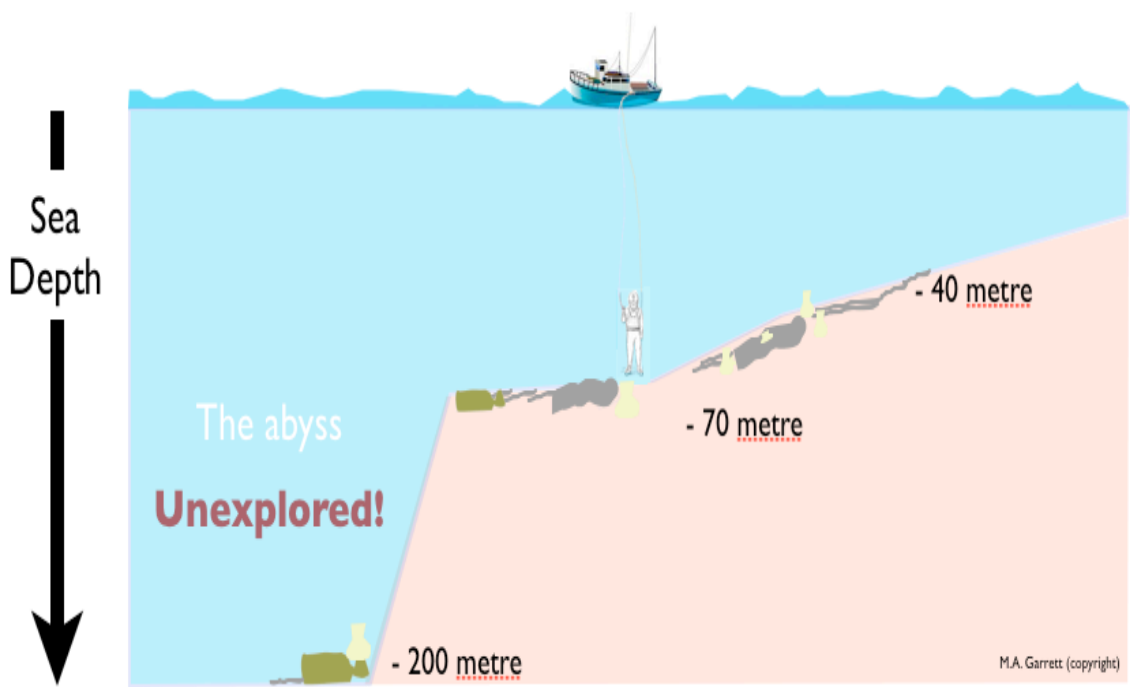

*Figure 5: a rough sketch (not to scale) of the seabed where the Antikythera wreck is to be found. Part of the site lies unexplored due to its depth. A systematic and scientific survey of the complete site and the surrounding area is now possible.*

In addition, even since the dive of Cousteau and his team some almost 40 years ago, diving technology has progressed significantly with the use of AUVs (Autonomous Underwater Vehicles), metal detectors, and sonar imaging equipment. This begs the question whether the Antikythera wreck should be revisited or not. Private discussions with Lefteris Tsavliris (see also these proceedings), a Greek diver who supported the Cousteau expedition suggested that revisiting the wreck should indeed be considered. Buried deep in the sediment, the hull of the ship is known to be largely intact - any cargo stored there might be very well preserved. We also know from the Archaeological Society's 1902 report, that several large boulders, lying on top of the wreck, were heaved out of the way and over the cliff before it was realised that these were actually enormous (but very corroded) statues [8]. A large statue of a horse was also lost when it slipped from its chains, as it was being winched onto a boat. The horse tumbled across the shelf and into the abyss (see Figure 5).

I raised the possibility of returning to the wreck during my presentation at the workshop, and was surprised by the obvious resistance that arose. For an astronomer, returning to an interesting astronomical find and subjecting it to all possible means of investigation, is natural and leads to a better understanding of the object. My impression is that for some Antikythera experts, the incomplete nature of the puzzle is part of the devices fascination. Nevertheless, while preparing this paper, I received news (John Reynolds, private communication) that the Wood Hole Oceanographic Institute of the United States were indeed about to begin surveying the site soon. This, in my opinion is very good news – who knows what this new expedition may reveal – more fragments of the mechanism, additional examples of similar devices, possibly additional wrecks of supporting vessels in the same area? We wait with interest and not a little anticipation for the first results to appear!